\documentclass[aps,prb,preprint,nofootinbib]{revtex4-1}

\usepackage{graphicx}
\usepackage{textcomp}
\usepackage{subfig}
\PassOptionsToPackage{colorlinks=true,linkcolor=black,citecolor=black,urlcolor=black}{hyperref}

\newcommand*\chem[1]{\ensuremath{\mathrm{#1}}}

\begin{document}


\title{Stoichiometry determination of chalcogenide superlattices by means of X-ray diffraction and its limits} 



\author{Henning Hollermann*}
\affiliation{I. Physikalisches Institut (IA), RWTH Aachen University, 52056 Aachen, Germany}
\author{Felix Rolf Lutz Lange*}
\affiliation{I. Physikalisches Institut (IA), RWTH Aachen University, 52056 Aachen, Germany}
\affiliation{JARA-FIT Institute Green-IT, RWTH Aachen University and Forschungszentrum J\"ulich, 52056 Aachen, Germany}
\author{Stefan Jakobs}
\author{Peter Kerres}
\affiliation{I. Physikalisches Institut (IA), RWTH Aachen University, 52056 Aachen, Germany}
\author{Matthias Wuttig}
\affiliation{I. Physikalisches Institut (IA), RWTH Aachen University, 52056 Aachen, Germany}
\affiliation{JARA-FIT Institute Green-IT, RWTH Aachen University and Forschungszentrum J\"ulich, 52056 Aachen, Germany}

\email{wuttig@physik.rwth-aachen.de}


\date{February 27, 2019}

\begin{abstract}
In this paper we explore the potential of stoichiometry determination for chalcogenide superlattices, promising candidates for next-generation phase-change memory, via X-ray diffraction. To this end, a set of epitaxial \chem{GeTe}/\chem{Sb_2Te_3} superlattice samples with varying layer thicknesses is sputter-deposited. Kinematical scattering theory is employed to link the average composition with the diffraction features. The observed lattice constants of the superlattice reference unit cell follow Vegard’s law, enabling a straight-forward and non-destructive stoichiometry determination 
\end{abstract}


\maketitle 

\section{Introduction}
\chem{\chem{GeTe}}/\chem{Sb_2Te_3} chalcogenide superlattices (CSL, also referred to as interfacial phase change memory (iPCM)) have attracted significant interest for next generation data storage. This interest arises from enhanced switching speeds, improved endurance as well as reduced power consumption compared to conventional PCRAMs. \cite{simpson_interfacial_2011} Beyond their application relevance, these superlattices (SLs) possess fascinating physical properties, such as topologically protected surface states,\cite{sa_topological_2012,tominaga_ferroelectric_2014} which renders CSLs interesting also for fundamental research. Indeed, superlattices in general are a very active area of research and SL-based applications like e.g. solid-state lasers \cite{nakamura_characteristics_1996} or thermoelectrics\cite{dresselhaus_new_2007} have become an integral part of modern technology. The key to reliable and predictable material and device performance, however, is based on a precise control of the SLs structure. A thorough structural characterization is thus mandatory to develop successful growth recipes and benchmark the quality of the SL structure.
Compared with doped semiconductors and alloys, SLs provide an additional level of complexity. While the stoichiometry of an alloy can be characterized by a single parameter, a SL requires two additional parameters to be defined precisely, namely the layer thicknesses of the two constituting materials (cf. Figure\,\ref{fig:1}(b)). Owing to this complexity, SLs span a huge parameter space to alter physical properties via tuning knobs such as the average composition of the SL, its bilayer thickness and the resulting interface density as well as the number of repetitions. Since it has been shown that intermixing is always present in these systems, there also have been studies to understand and hinder this tendency.\cite{momand_interface_2015,momand_atomic_2016,cecchi_improved_2017}
In the case of CSLs, the structural characterization mainly rests on studies using (scanning) transmission electron microscopy in conjunction with energy-dispersive X-ray spectroscopy ((S)TEM/EDX) as well as X-ray diffraction (XRD) methods. The combination of TEM and EDX is powerful to access all structural parameters on a single lamella. Studies utilizing these two techniques have already provided important information on the local atomic arrangement at \chem{\chem{GeTe}}/\chem{Sb_2Te_3} interfaces.\cite{momand_interface_2015,momand_atomic_2016,lotnyk_van_2018} TEM is unparalleled at the nanoscale, yet it suffers from time-consuming sample preparation and destructivity. EDX and XRD on the other hand are fast and non-destructive techniques to provide non-local chemical and structural information respectively without the need of any prior preparation. Nevertheless, XRD is mainly used to verify the SL character by the observation of peaks of larger intensity, that are surrounded by “satellite peaks” whose spacing can be translated to the unit cell size of the superlattice. \cite{speriosu_xray_1984,fullerton_structural_1992}
In the present work we show that the analysis of these characteristic features can be used to determine the average stoichiometry of any given CSL. The model relies on kinematical scattering theory and will be shown to be equivalent to a description in terms of Vegard’s law \cite{vegard_konstitution_1921} which assumes a linear relationship between the lattice size and the alloy composition. It has already been shown that GeSbTe alloys indeed follow Vegards Law. \cite{karpinsky_x-ray_1998} For SL structures, however, such a linear dependence is not obvious since both parent materials are subject to stress and strain. Thus an elastic response obeying the elastic constants of the parent materials is conceivable. 

In fact, literature provides examples of both cases, systems following\cite{madan_x-ray_1997,brunner_growth_1996,liang_effects_2006} and systems deviating from\cite{ishida_properties_1986,nikulin_model-independent_1996,lewis_structure_1999,monfroy_molecular_1986,kasper_test_1995,dismukes_lattice_1964} Vegard's law. In the case of III-V SLs, Vegard's law is frequently employed to estimate the stoichiometry of the SL,\cite{blefeld_preparation_1983,joncour_xray_1985,ng_molecular_2000} while for systems like CdTe/ZnTe,\cite{monfroy_molecular_1986} PbTe/EuTe\cite{ishida_properties_1986} and TiAlN/CrN\cite{lewis_structure_1999} but also for non-SL Si-Ge \cite{kasper_test_1995,dismukes_lattice_1964} deviations are reported. In the case of SLs which incorporate \chem{V_2IV_3} compounds such as \chem{Sb_2Te_3} the situation is potentially even more complex as these materials are composed of van-der-Waals-like gaps featuring a weak coupling of adjacent building blocks.\cite{wang_2d_2018,vermeulen_strain_2018} Hence, the applicability of Vegard’s law is not guaranteed. But only recently a linear dependence of the in-plane lattice parameter on the composition of natural \chem{Bi_nSe_m} superlattices was found by XRD simulations.\cite{springholz_structural_2018}

In this paper we will present a more practical approach, since SLs of \chem{GeTe} and \chem{Sb_2Te_3} at present is the most promising combination when it comes to iPCM applications and especially for industry, straight-forward methods are crucially relevant. In the final section the limits of the method are explored and possible error sources identified.

\section{Stoichiometry from the diffraction perspective}
\begin{figure}
\centering
\subfloat[]{\includegraphics[width=0.68\columnwidth]{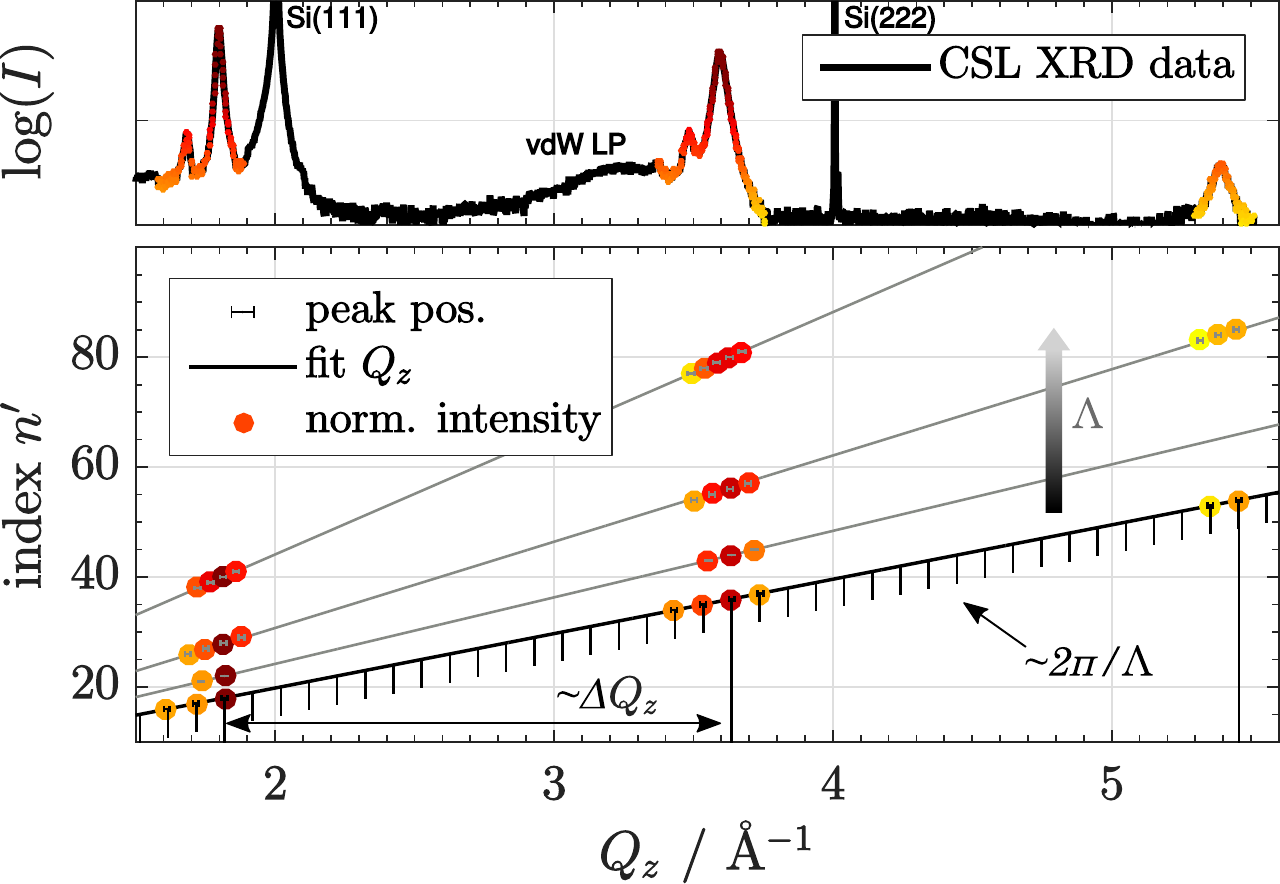}}
\subfloat[]{\includegraphics[width=0.3\columnwidth]{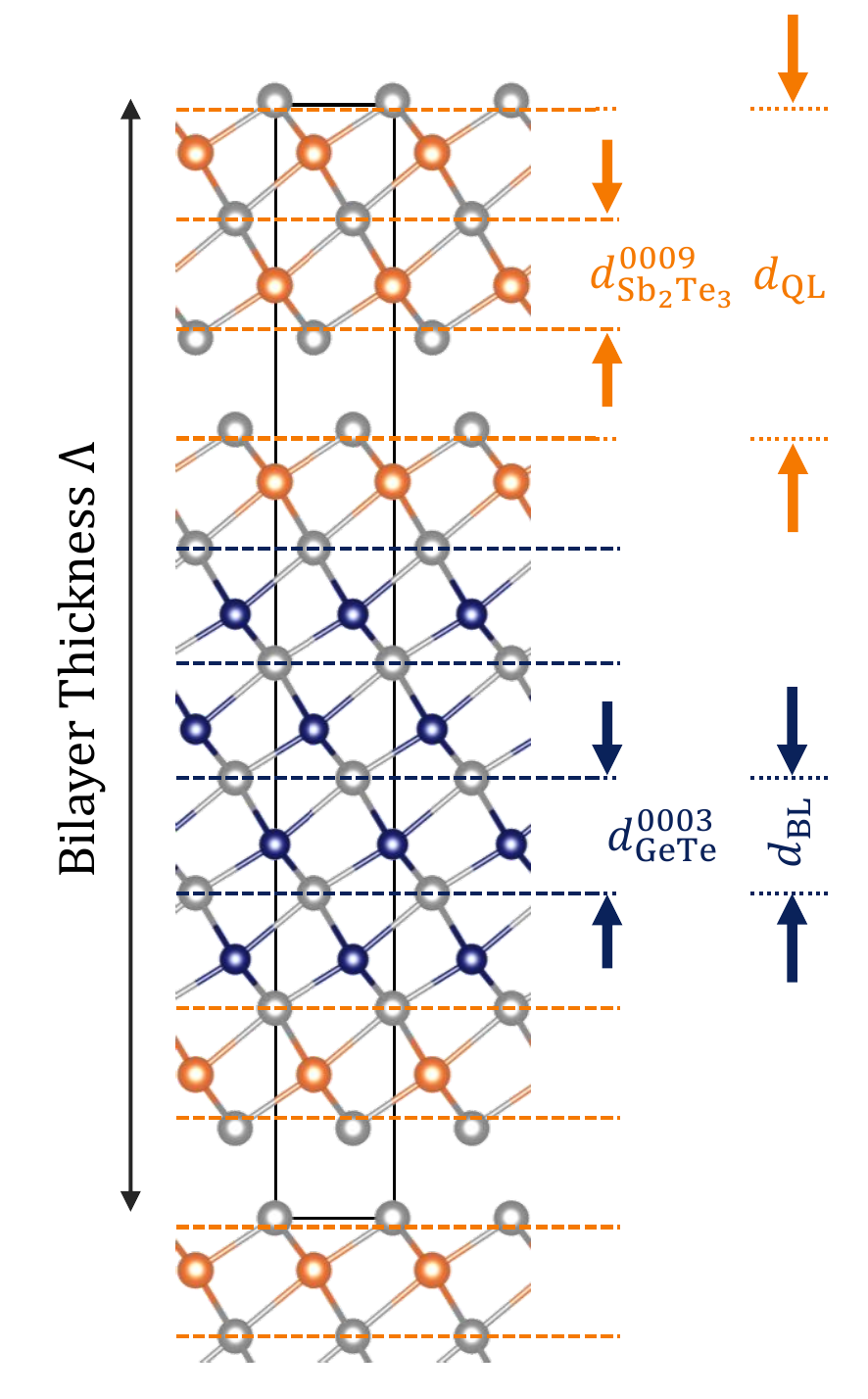}}
 \caption{(a) In the upper panel a typical diffraction pattern of a superlattice is shown. It is apparent that only a small number of closely spaced diffraction peaks stemming from the SL supercell are observed. This is due to the variation of supercell size that happens for all CSL samples during deposition. In the lower panel the positions of these superlattice peaks are plotted for a number of superlattices against the peak index, showing that each set of peaks belongs to one single plane family, namely the respective supercell. The different samples feature different supercell sizes, but they share the same composition. All four superlattices show the strongest diffraction peaks at similar positions in reciprocal space (red points). Their spacing translates to the mean Te-Te layer distance in the superlattice (reference lattice), which is a direct measure for composition. The asymmetric distribution of intensity within each peak group can also be attributed to the underlying structure factor of the supercell. (b) Depicted is model representation of a SL supercell as a simple epitaxial stacking of \chem{GeTe} bilayers (BLs) and \chem{Sb_2Te_3} quintuple layers (QLs) projected along [001]. Also given are $d_\chem{Sb_2Te_3}^{0009}$ and $d_{\chem{\chem{GeTe}}}^{0003}$ - the mean Te layer distances of either material. In the case of \chem{Sb_2Te_3}, we have van-der-Waals like gaps and therefore the (0009)-planes do not coincide with atom positions. The full spectra of the lower panel of (a) can be found in the supplement.
\label{fig:1}}%
 \end{figure}

A superlattice, as a repeated epitaxial stacking of two (or more) materials, is described by its own unique unit cell, called supercell. The out-of-plane periodicity of the SL is given by the sum of the thicknesses of the parent materials and is referred to as the bilayer thickness $\Lambda$. Therefore, the set of possible diffraction maxima on the specular crystal truncation rod is $Q_z^{(n^\prime)}=4\pi \frac{\sin\Theta}{\lambda}=n^\prime \frac{2\pi}{\Lambda}$. Figure\,\ref{fig:1} shows the diffraction pattern of such a superlattice. All sharp diffraction peaks can be explained as integer multiples of the primitive reciprocal lattice $2\pi/\Lambda$, directly related to the bilayer thickness. Figure\,\ref{fig:1}(a) shows the peak positions extracted from XRD data for several such superlattice samples with different $\Lambda$ plotted against the peak index $n^\prime$. From the slope of the linear fit the supercell size can be determined (see suppl. material for details). Yet, it is striking that for this system only a few of these peaks appear as groups around distinct positions in $Q_z$. Although certain features like the broad peak at around $Q_z=3.1 \textup{\AA}^{-1}$ have already been discussed previously \cite{bragaglia_metal_2016,wang_intermixing_2016,zallo_modulation_2017}, a general description of the pattern in terms of kinetic diffraction theory is still lacking. It will also help to clarify the effect of the SL stoichiometry on the diffraction patterns observed.

As we pointed out, the crystal lattice $Q_z^{(n^\prime)}$ is defined by the supercell size $\Lambda$. In the following, we will discuss the intensity variation in terms of a)  the structure factor of the supercell and b) defects and disorder in real samples.

\begin{figure}[htb]
\includegraphics[width=0.7\columnwidth]{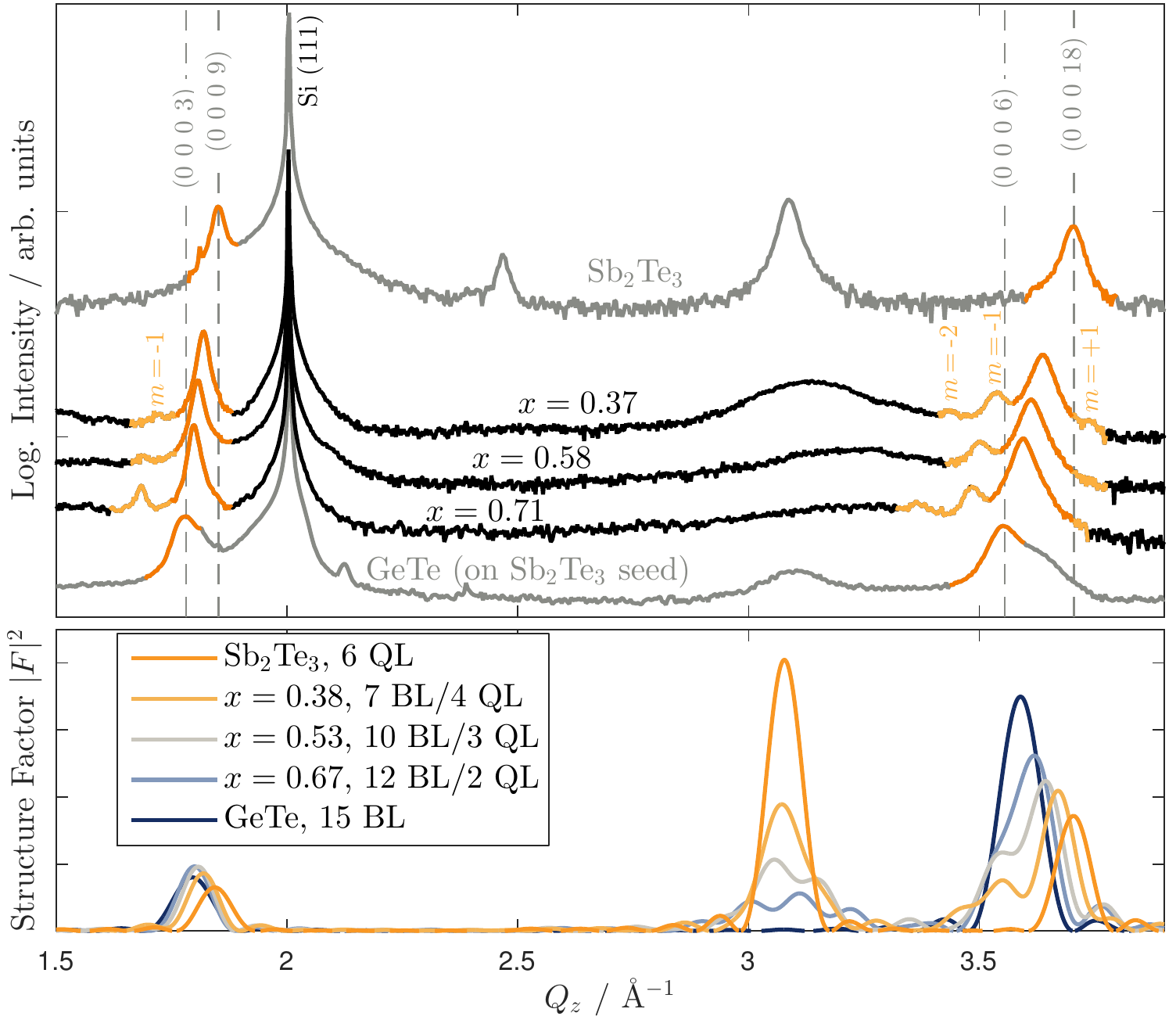}
 \caption{Overview of diffraction patterns for \chem{GeTe}/\chem{Sb_2Te_3} superlattices of similar supercell size $\Lambda$ but different stoichiometry together with \chem{GeTe} and \chem{Sb_2Te_3} reference samples. Both parent materials are indexed using the hexagonal axes. Their structure factor only allows for (0003L) peaks which coincide in regions where we also find the peaks of the superlattice. They  appear as groups around the mean Te-Te layer distance, which are usually labeled as superlattice peaks (orange) and neighboring satellites (light orange). With decreasing \chem{GeTe} content $x$ and therefore a shift to more \chem{Sb_2Te_3}-rich SLs, the peak groups shift to larger $Q_z$, since the mean Te-Te distance decreases. This is also visible in the structure factors (lower panel). The peakshift can be used to determine the stoichiometry from diffraction data. With increasing \chem{Sb_2Te_3} content, we see the so called van-der-Waals-layer peak emerging around $Q_z=3.1\,\textup{\AA}^{-1}$, which is due to the large contribution of \chem{Sb_2Te_3} to the structure factor at this position. Note that the \chem{GeTe} reference sample was grown on an \chem{Sb_2Te_3} seed to improve the texture.}%
\label{fig:2}
 \end{figure}

Figure\,\ref{fig:2} (upper panel) displays diffraction patterns of three SL samples with similar $\Lambda$ but different composition $x$, together with \chem{Sb_2Te_3} and \chem{GeTe} reference samples. The lower panel shows the corresponding calculated structure factors (c.f. supplemental material). The rhombohedral \chem{GeTe} with the Ge-Te bi-layers (BL) and the trigonal \chem{Sb_2Te_3} with Te-Sb-Te-Sb-Te quintuple layers (QL) as relevant motifs both contribute to the total structure factor. Therefore the main contributions change with stoichiometry, and so does the position of the peak groups in the experimental data, because the structure factor is responsible for the intensity modulation of the crystal lattice peaks at $Q_z^{(n^\prime)}$.

The real-space equivalent of the spacing of the peak groups is the mean Te-Te layer distance of the superlattice. Both our constituents have a common sublattice of Tellurium planes. However, the mean Te-plane distance of each material is different, mainly because the \chem{Sb_2Te_3} structure consist of QL that are coupled across van-der-Waals like gaps that have a smaller Te-Te distance as compared to the Te-Ge-Te distance within the BL (c.f. Figure\,\ref{fig:1}(b)). With an increase of the \chem{GeTe} content, and thus a reduction of the number of van-der-Waals like gaps, we therefore increase the mean Te-Te distance and consequentially the peak groups shift to smaller $Q_z$ values. We hence use the mean Te-Te distance to model the structure-stoichiometry relation of our superlattice samples.

Also in Figure\,\ref{fig:2} one can observe that for larger \chem{Sb_2Te_3} content a broad feature emerges at around $Q_z=3.1 \textup{\AA}^{-1}$ which was called vacancy layer (VL) peak by some authors \cite{bragaglia_metal_2016,wang_intermixing_2016}. Our structure factor plot shows, that its intensity is due to the large contribution of \chem{Sb_2Te_3} (QL) at this position, also responsible for the pronounced intensity of \chem{Sb_2Te_3}(00015) (c.f. suppl. material). The fact, that we do not observe sharp satellite peaks in these regions of reciprocal space is mainly due to imperfections of the superlattice, introducing disorder.

The significance of disorder becomes evident upon comparison to similar samples grown by molecular beam epitaxy (MBE) or when sputtering them on muscovite mica substrates. Such films, which can be produced with higher crystal quality, show a few more sharp peaks over slightly wider regions of reciprocal space. This implies that film imperfections cause a significant number of diffraction peaks to vanish. From TEM it is known that in a CSL not only one single supercell size is favored, but various \chem{GeTe} and \chem{Sb_2Te_3} layer thicknesses and also layers with a GeSbTe stacking exist within the CSL stack. \cite{momand_dynamic_2017} During deposition variations by one or more BL or QL are highly likely. Since the scattering volume is much larger than the coherence length of the X-rays, the measured intensity is an average of all the slightly differently stacked grains within that volume (configurational average). Consequentially, only peak groups with sharp satellites (indicative for the mean lattice of Te-planes) persist, because away from these regions in reciprocal space, the lattices of different supercell sizes dephase (cf. suppl. material). This finding and also our discussion of  the VL peak above are in line with recent results \cite{cecchi_improved_2017} where it was shown that an exchange of \chem{GeTe} with GST dramatically increases the visibility of satellite peaks all over the $Q_z$ axis.

In the literature the most intense peaks at $Q_z^{(m)} \approx m \cdot 1.82 \textup{\AA}^{-1}$ are referred to as superlattice peaks (highlighted in orange in Figure\,\ref{fig:2}) which are flanked by a number of so called satellite peaks (lighter orange). As explained above, this naming scheme is a convenient description only -- all peaks belong to the same supercell plane family.

However, these labels stem from the reference lattice concept,\cite{holy_high-resolution_1999,pietsch_high-resolution_2004-1} which is often employed to describe the diffraction pattern of multilayer structures. \cite{khan_structural_1983,madan_x-ray_1997,fullerton_structural_1992} We will now show how the concept can be used to deduce the chemical composition of a given SL sample from XRD-data. In a pseudomorphic superlattice, the supercell size $\Lambda$ is composed of a discrete number of lattice planes of the parent materials
\begin{equation}
	\Lambda=n_1 d_1+n_2 d_2
    \label{eq:lambda_def}
\end{equation}
with $d_1$ and $d_2$ being lattice plane spacings of the constituent materials. For the SL discussed here the plane distances $d_{\chem{\chem{GeTe}}}^{(0003)}$ and $d_{\chem{Sb_2Te_3}}^{(0009)}$ correspond to the mean Te-Te layer distances in either material and consequentially $n_1$ and $n_2$ are the number of those Te layers (c.f. Fig.\,\ref{fig:1}(b)). 
To complete the reference lattice description we use Eq.\,(\ref{eq:lambda_def}) to define a fraction $\bar d$ of \emph{the supercell} as the mean plane spacing (Te-Te layer spacing)
\begin{equation}
	\bar d := \frac{\Lambda}{n_1+n_2}
    \label{eq:dbar_lambda_def}
\end{equation}
The diffraction peaks in terms of the reference lattice are therefore indexed by a modified Bragg’s law as\cite{fullerton_structural_1992,pietsch_high-resolution_2004-1}
\begin{equation}
	Q_z^{(n^\prime)} = n^\prime  \frac{2\pi}{\Lambda} \Leftrightarrow Q_z^{(m,n)} = m \cdot \frac{2\pi}{\bar d} \pm n\cdot \frac{2 \pi}{\Lambda}
    \label{eq:braggs_law}
\end{equation}
with $m,n=0,1,2...$ and $n^\prime=m(n_1+n_2)+n$. Peaks labeled by m are the superlattice peaks belonging to the reference lattice and peaks indexed by $n$ are the superlattice satellites (c.f. Fig.\,\ref{fig:2}).

Since the stoichiometry determination relies on the difference in lattice plane distances between $d_1$ and $d_2$, the average Te-Te layer spacing $\bar d$ can be used to determine the compositional fraction $x =\frac{n_1}{n_1+n_2}$ of the constituent materials as follows
\begin{equation}
	\bar d = d_1 x + (1-x) d_2
    \label{eq:dbar_formula}
\end{equation}

It may be noted that the absolute values of $\bar d$ and $x$ depend on the choice of $d_1$ and $d_2$ that describe the supercell (c.f. Eq.\,(\ref{eq:lambda_def})).

With Eq.\,(\ref{eq:dbar_formula}) it becomes evident that $\bar d$ scales monotonously and linearly with the amount of either constituent material, resembling Vegard's law for alloys as described above. Therefore, a successful description in terms of Eq.\,(\ref{eq:dbar_formula}) will verify the applicability of Vegard's law.

If Eqs.\,(\ref{eq:braggs_law}) and (\ref{eq:dbar_formula}) are combined the compositional ratio of a SL can be obtained from a diffraction experiment as
\begin{equation}
	x =\frac{d_2 -2\pi/\Delta Q_z}{d_2-d_1},
    \label{eq:eta_formula}
\end{equation}
where $\Delta Q_z=Q_z^{(m+1,0)}-Q_z^{(m,0)}$ is the distance of two subsequent SL peaks. In the following, after giving information about the methods used, we will present and discuss the experimental results on the stoichiometry dependence of $\bar d$. 

\section{Methods}

To validate the model a series of \chem{\chem{GeTe}}/\chem{Sb_2Te_3} SLs are sputter deposited from stoichiometric targets with a purity of 99.99 at.\% on Si(111)-H substrates. The substrates were dipped in 1\% aqueous hydrofluoric acid for 30\,s and transferred to the vacuum chamber immediately after for further processing. Deposition temperatures ranging from 180$^\circ$C to 210$^\circ$C, 35\,W DC power and Argon gas flux of 35\,sccm were used. By variation of the deposition times used for both materials, the thicknesses $t_{\chem{Sb_2Te_3}}$ and $t_{\chem{\chem{GeTe}}}$ are varied. The average plane spacing $\bar d$ is obtained for each sample by X-ray diffraction using a Bruker D8 Discover setup equipped with a Goebel mirror and a two-bounce Ge(220) asymmetric channel-cut monochromator. The compositional ratio $x$ is measured by energy-dispersive X-ray spectroscopy with an FEI Helios 650 NanoLab system (electron beam: 10 keV, 0.4 nA, scan-area: $200\,\textup{um}\times300\,\textup{um}$) and the Oxford Instruments AZtec software (version 2.1) for data recording and analysis. Additional information on sample characterization is detailed in the supplementary information.

\section{Results}

The lattice mismatch of the constituent materials amounts to 2.5\% for the (0001) orientation. As it has been demonstrated in previous work,\cite{momand_atomic_2016} for this growth direction and elevated temperatures we expect quasi-van-der-Waals epitaxy of the two materials and thus the formation of a CSL (c.f. suppl. material).
As motivated above, our SL samples are conveniently described by a reference lattice with $\bar d \approx 3.45\,\textup{\AA}$, defined by the plane distances $d_1=d_{\chem{\chem{GeTe}}}^{(0003)}$ and $d_2=d_{\chem{Sb_2Te_3}}^{(0009)}$ in Eq.\,(\ref{eq:lambda_def}) (c.f. Fig.\,\ref{fig:1}(b)). Hence we index the SL peaks in Fig.\,\ref{fig:2} as (0 0 0 $m$).
In the limit of Vegard's law, also the in-plane lattice constant a can be expressed similarly to Eq. (\ref{eq:dbar_formula}) as
\begin{equation}
	a=a_{\chem{\chem{GeTe}}} x + (1-x) a_{\chem{Sb_2Te_3}},
    \label{eq:a_formula}
\end{equation}
In order to experimentally test Eqs.\,(\ref{eq:dbar_formula}) and (\ref{eq:a_formula}), a set of plane families has been measured by XRD, namely $\{0 0 0 1\}$, $\{1 1 \bar 2 1\}$, $\{1 1 \bar 2 2\}$ and $\{1 1 \bar 2 3\}$. Their peak center was determined by subsequent $\varphi$, $\psi$, $\omega$, $\Theta$ and $2\Theta$/$\omega$ measurements in bisecting geometry. The $a$- and $\bar d$-axes were subsequently obtained by a least-square algorithm. In addition, the $c$-axis was also determined by $\Theta$-$2\Theta$ scans (along $Q_z$) for a larger set of samples. Here, a linear regression of $Q_z^{(m,0)}$ with diffraction order $m$ was used to obtain $\bar d$.
The corresponding EDX data was analyzed assuming a stoichiometric (\chem{\chem{GeTe}})$_x$(\chem{Sb_2Te_3})$_{1-x}$ SL. The EDX instrument was tested comparing the results at 10\,keV with results from wavelength-dispersive X-ray spectroscopy (WDX), TEM and atom probe tomography (APM). As the results are in very good agreement, we estimate the error value to be 1\,at.\%. All three combinations of ratios (Ge/Sb, Ge/Te and Sb/Te) were used to calculate $x$. The final value is a weighted average leading to uncertainties in the \chem{GeTe} content of about $\sigma_x\sim 0.02$.
Fig.\,\ref{fig:3} shows the result of this investigation. As can be seen, Eqs.\,(\ref{eq:dbar_formula}) and (\ref{eq:a_formula}) (dashed-dotted lines) nicely reproduce the experimental data. The change in unit cell thus follows Vegard's law. The model was fitted with $a_{\chem{\chem{GeTe}}}$ and $a_{\chem{Sb_2Te_3}}$ or $c_{\chem{\chem{GeTe}}}$ and $c_{\chem{Sb_2Te_3}}$ as free parameters, respectively. Their values $a_{\chem{Sb_2Te_3}}=4.272\,\textup{\AA}$, $c_{\chem{Sb_2Te_3}} = 9 \cdot d_{\chem{Sb_2Te_3}}^{(0009)}=30.589\,\textup{\AA}$, $a_{\chem{\chem{GeTe}}}=4.180\,\textup{\AA}$ and $c_{\chem{\chem{GeTe}}}=3\cdot d_{\chem{\chem{GeTe}}}^{(0003)}=10.632\,\textup{\AA}$ agree well with data reported in literature for bulk samples $a_{\chem{Sb_2Te_3}}=4.264(1)\,\textup{\AA}$, $c_{\chem{Sb_2Te_3}}=30.458(5)\,\textup{\AA}$,\cite{anderson_refinement_1974} $a_{\chem{\chem{GeTe}}}=4.156(3)\,\textup{\AA}$ and $c_{\chem{\chem{GeTe}}}=10.663(5)\,\textup{\AA}$.\cite{pereira_lattice_2013} Therefore the model can be used to estimate the stoichiometry by obtaining either the lattice parameter $a$ or $\bar d$ of the reference lattice and use Eqs.\,(\ref{eq:eta_formula}) to calculate $x$. 

It should be noted, that with knowledge of $x$ and $\Lambda$ (obtained from the SL peak position and the satellite spacing) also the layer thicknesses $t_{\chem{\chem{GeTe}}}$ and $t_{\chem{Sb_2Te_3}}$ of the parent materials can be deduced that comprise $\Lambda$.

\section{Discussion}

The last part of this paper illuminates potential error sources and estimates the error of this method depending on a) the stoichiometry ratio $x$, b) the bilayer thickness $\Lambda$, and c) the interface density $\rho\sim\Lambda^{-1}$.


Eqs.\,(\ref{eq:dbar_formula}) and (\ref{eq:a_formula}) are based on a precise determination of the peak center of the SL reference peaks. This becomes challenging if the bilayer thickness $\Lambda$ becomes large enough for the peak areas of the zeroth order Bragg peak and its satellites to overlap. 
In this case, a proper deconvolution of peaks is mandatory to obtain $Q_z^{(m,0)}$. Assuming a full-width at half-maximum of $\beta=0.02\,\textup{\AA}^{-1}$ of the SL peaks and satellites, the maximum bilayer thickness without significant overlap is estimated to be $\Lambda_\textsubscript{max}=\frac{2\pi}{2\beta}=157\,\textup{\AA}$.
This value amounts to roughly three times the bilayer thickness reported by Simpson et al.\cite{simpson_interfacial_2011} for successful device operation ($\Lambda\approx 47 \textup{\AA}$). Yet, this estimation relies on the structural quality of the CSL as $\beta$ broadens with the defect density, the vertical coherence length and stress/strain in the system.

The model presented only considers the simple case of two plane spacings.
As shown by Momand et al.,\cite{momand_interface_2015,momand_atomic_2016} CSLs deposited at elevated temperatures tend to intermix at the interfaces, thus forming building blocks of thermodynamically stable (\chem{\chem{GeTe}})$_x$(\chem{Sb_2Te_3})$_{1-x}$ alloys (GST). 
It has been also shown that strain plays a role in the epitaxy of CSL structures\cite{wang_2d_2018,vermeulen_strain_2018}.
This adds further complexity to Eq.\,(\ref{eq:lambda_def}) since in principle a third plane spacing would have to be included.
In this sense the model presented is a simplification that may lead to inaccuracies describing the data. As the effect of intermixing and also the strain-gradients discussed in \cite{wang_2d_2018} are expected to scale with the interface density $\rho=\Lambda^{-1}$, several CSL have been sputter deposited with the same composition $x$, but varying bilayer thicknesses $\Lambda$ ranging from 20 to 140\,\textup{\AA}.

\begin{figure}%
\includegraphics[width=0.7\columnwidth]{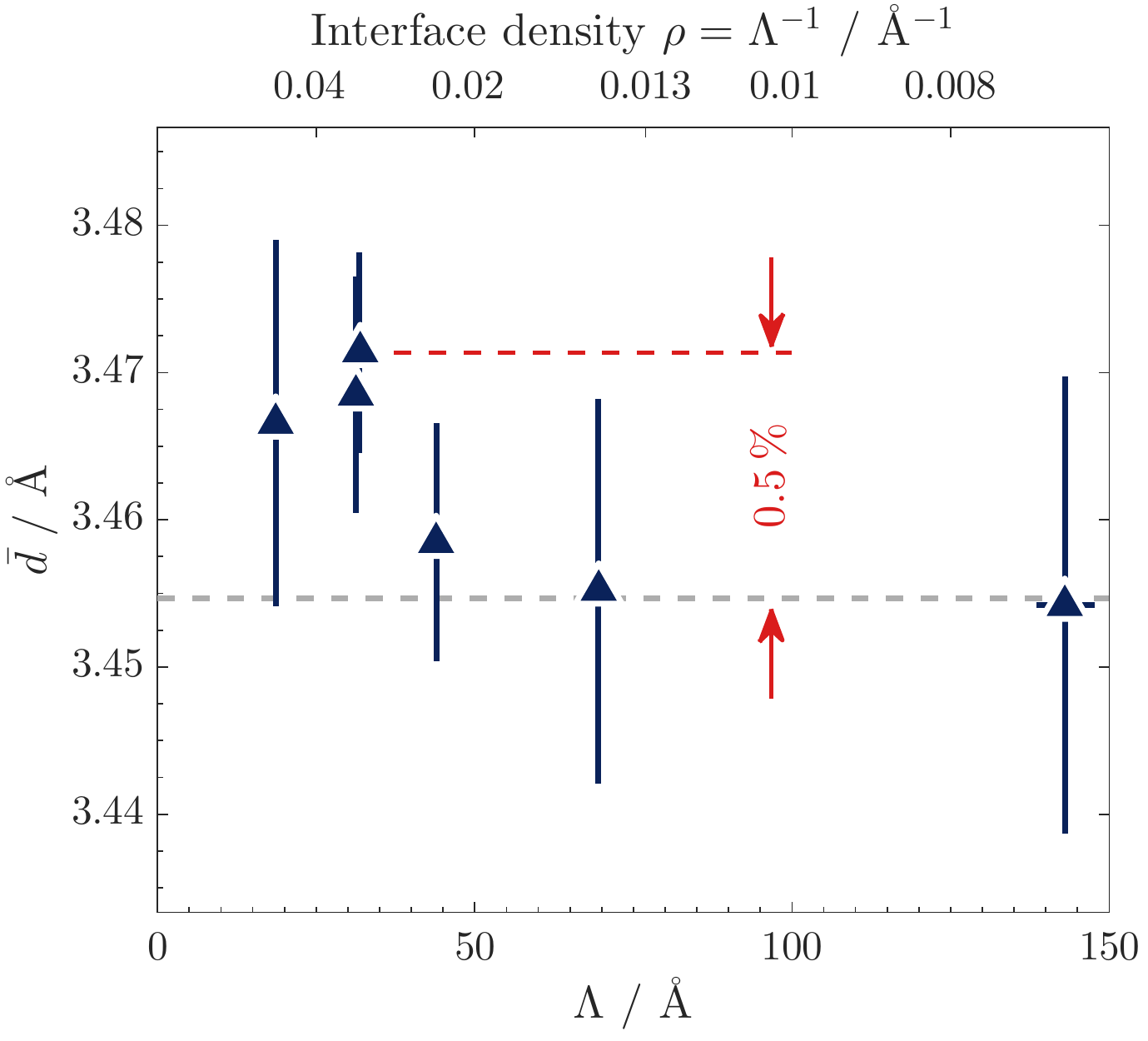}%
\caption{Change of reference length $\bar d$ with interface density $\rho$. Here, the compositional ratio is kept constant ($\eta=1.8\pm0.1$). $\bar d$ increasingly deviates from Eq. (\ref{eq:dbar_formula}) when approaching large interface densities (smaller values of $\Lambda$). The overall descrepancy, however, is less than 0.5\%, thus rendering $\bar d$ still a good measure.}
\label{fig:4}%
\end{figure}

%
%
%
%
As can be seen from the results shown in Fig.\,\ref{fig:4}, for high interface densities (small values of $\Lambda < 40\,\textup{\AA}$) the $\bar d$-axis of the reference lattice deviates from the proposed value given by Eq.\,(\ref{eq:dbar_formula}) (dashed-dotted line in Fig.\,\ref{fig:4}) representing an uncertainty in $x$ of about 12\%. For medium interface densities, the deviation however, stays below 5\%, thus proving that the model still provides a good estimate. This indicates that the strain which is present in the SL (depending both on $x$ and $\rho$) is below the detection limit of our method. 

\begin{figure}
\includegraphics[width=0.7\columnwidth]{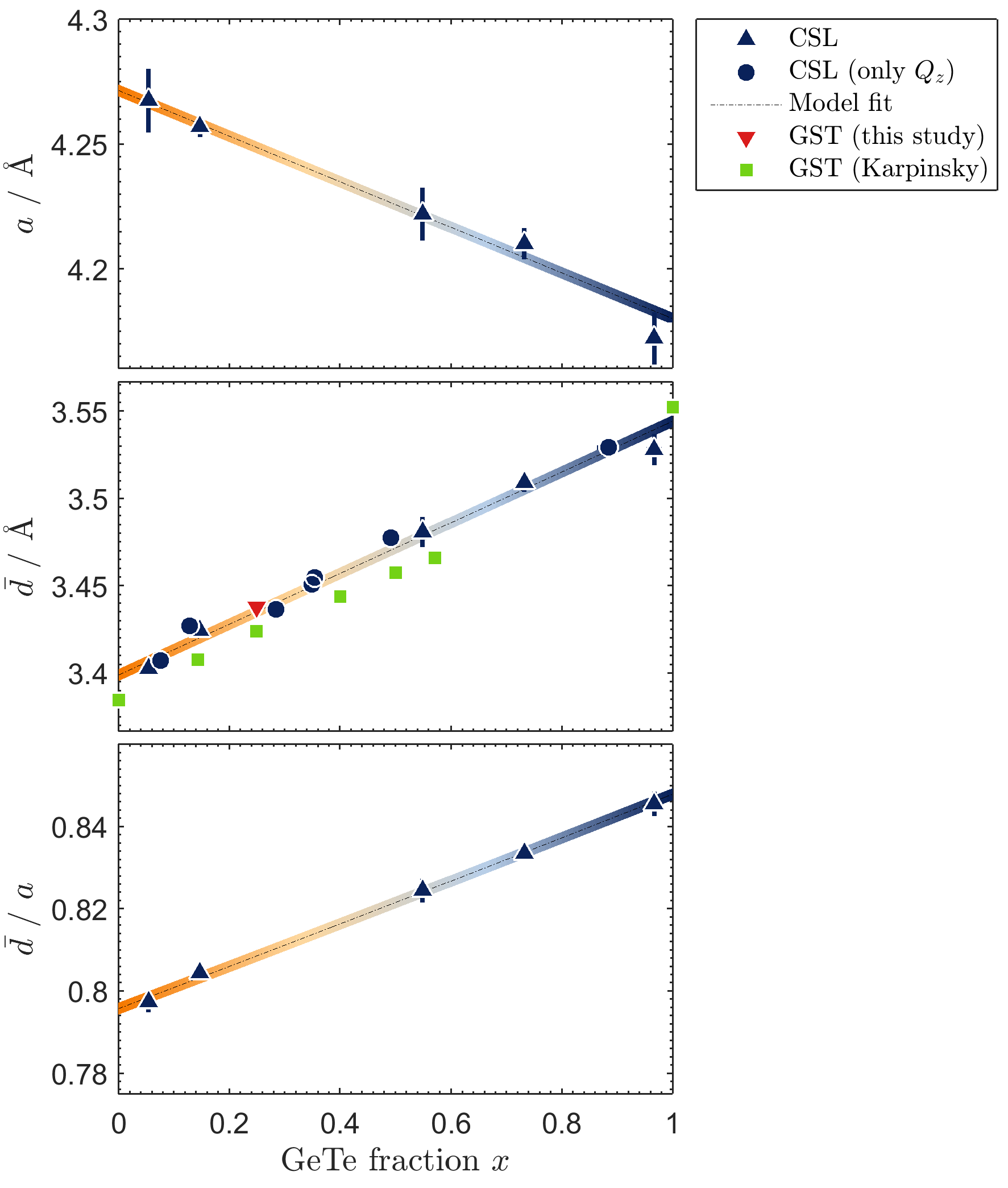}
 \caption{Evolution of the SL reference lattice unit cell with composition. The data can be found in the supplemental material. Both parameters, $\bar d$ and a follow \ref{eq:dbar_formula} and (\ref{eq:a_formula}), thus obeying Vegard's law. Moreover, it is found that CSLs follow the same generic behavior as their related $(\chem{\chem{GeTe}})_x(\chem{Sb_2Te_3})_{1-x}$ alloys (GSTs). The corresponding values for bulk samples are taken from Karpinsky et al..\cite{karpinsky_x-ray_1998} The GST film (red triangle) refers to a \chem{GeSb_2Te_4} thin-film, deposited at 300°C.
\label{fig:3}}%
 \end{figure}

For high-interface densities we assume that due to intermixing the CSL structure dissolves into the stable rhombohedral GST structure which -- unlike the meta-stable cubic phase\cite{matsunaga_single_2006} -- also follows Vegard's law as reported by Karpinsky et al..\cite{karpinsky_x-ray_1998} Their data points are also displayed in Fig.\,\ref{fig:3} (green squares). 
Using first-principle calculations, the experimental values were corroborated by Da Silva et al. for both, the $a$ and $c$ axes of bulk GST.\cite{da_silva_insights_2008} Interestingly, the authors also understand the GST structure as a superlattice-like-stacking of \chem{Sb_2Te_3} and \chem{\chem{GeTe}}. Clearly, a CSL unit cell made of a single \chem{\chem{GeTe}} bilayer intercalated into \chem{Sb_2Te_3} quintuples results in a stable GST structure. To verify this generic behavior, a \chem{GeSb_2Te_4} thin-film was sputter-deposited at 300$^\circ$C from a stoichiometric target. As can be seen in Fig.\,\ref{fig:3} (red triangle), also this sample is nicely described by the model presented above.

In summary, the limits of stoichiometry determination via X-ray diffraction for
\chem{\chem{GeTe}}/ \chem{Sb_2Te_3} superlattices have been explored. The measured change in lattice constants obeys Vegard's law. This insight can be used to estimate the stoichiometry of any SL by a straight-forward evaluation of diffraction peaks. Intermixing at the \chem{\chem{GeTe}}/\chem{Sb_2Te_3} interface does not affect the reference unit cell noticeably. This method thus provides a fast complementary method to TEM/EDX for stoichiometry determination – even for CSL structures in complex device structures. The results obtained here can be extended to related SLs systems composed of (IV-VI):(V\textsubscript{2}VI\textsubscript{3}) materials or other $p$-bonded chalcogenides.

\begin{acknowledgments}
The financial support by the DFG through SFB 917 is gratefully acknowledged. This work was also partially funded by the PASTRY project (GA 317746) within the FP7 of the EU.
\end{acknowledgments}

\section*{Contributions}
Henning Hollermann and Felix R. L. Lange contributed equally to this work.

\bibliography{./eta-CSL}

\clearpage

%
%

%


\end{document}